\documentclass[11pt]{article}
\usepackage{hyperref}
\pdfoutput=1
\begin{document}
\title{A Pitching Foil with a Flexible Flap Creates an Orderly Jet}
\author{Sachin Y. Shinde and Jaywant H. Arakeri \\
\\\vspace{6pt} Department of Mechanical Engineering, \\ Indian Institute of Science, Bangalore-560012, India}
\maketitle

          The wing-flexibility of flying birds and insects is one
          important reason, among the others, for high lift generation
          during hovering [1]. Inspired by this observation, we
          experimentally studied the effect of chordwise flexibility
          on the flow generated by flapping foil in the quiescent
          fluid.

          To mimic wing-flexibility, a flexible flap with negligible
          mass and stiffness is attached at the trailing edge of
          NACA0015 airfoil that oscillates sinusoidally at a fixed
          location in quiescent water in a glass tank. The airfoil is
          confined between the end-plates to ensure two-dimensionality
          of the flow. The airfoil model is oscillated by a servo motor
          about a hinge point at 30\% chord from the leading edge. The
          dimensions of the airfoil are: chord 40 mm and span 100
          mm. The flap length is 75\% of the rigid chord length; the
          flap is 100 mm in spanwise direction. The flap is made from
          0.05 mm thick Polythene sheet; its Young's modulus (E) is
          $3.02 \times 10^{8}$ $\mathrm{N/m^2}$ and flexural rigidity
          (EI) is $3.15 \times 10^{-7}$ $\mathrm{Nm^2}$.  We visualize
          the flow with Polystyrene particles and measure velocities
          using particle image velocimetry (PIV).  The parameters that
          were varied were frequency and amplitude of oscillation. The flow
          is studied for 12 cases: three amplitudes of oscillation,
          $\pm 10^o$, $\pm 15^o$, $\pm 20^o$, and four frequencies for
          each amplitude, 1, 2, 3 and 4 Hz. The experiments were
          conducted with two airfoil models: one with flexible flap
          and the other without flap.

          We observed that sinusoidal pitching of the airfoil without
          flap i.e. with sharp, rigid trailing edge in still water
          produces a divergent, weak jet that meanders randomly about
          the mean-position, in all the 12 cases. On the contrary, the
          airfoil with flexible flap creates a narrow, coherent,
          non-meandering, undulating jet with the vortices staggered
          in the form of a `reverse Karman vortex street', for most of
          the cases studied.  The non-meandering, orderly jet produced
          by airfoil with flexible flap stays nearly along the
          mean-position line.  The flow generation mechanism is
          as follows.  The water is drawn-in by the rigid foil as well as
          flexible flap from the front and mainly from the sides
          towards the airfoil model and given momentum and energy in a
          highly directional manner to form the jet. The flap
          undergoes large deformations and plays very important role
          in keeping proper spacing among the vortices by shedding
          them at appropriate points and phases such that they are
          sustained far downstream.

          The major differences between the flows produced by the two
          airfoil models suggest that the flexibility of flap is
          crucial in the production of such an orderly jet. But, one
          important question arises i.e. does the airfoil with
          flexible flap always produce an orderly jet? It has been
          observed that the foil with flexible flap does not produce
          an orderly jet for the extreme cases in the parameter
          set. Effective stiffness of the flap reduces and it becomes
          more flexible with increase in amplitude and frequency of
          oscillation.  The flexible foil produces a jet that meanders
          about the mean-position line for amplitude $\pm 10^o$ and
          frequency 1 Hz where the flap is relatively stiff, and, it
          produces a jet that blooms and spreads in the downstream
          region for amplitude $\pm 20^o$ and frequency 4 Hz where the
          flap is effectively very flexible. Thus, the creation of narrow,
          coherent, non-meandering, orderly jet in quiescent fluid is
          possible only with appropriate chordwise flexibility of the
          flap.

          During hovering, the vortex rings shed by the animal take
          the form of a jet of air blowing down vertically below the
          animal. This jet has thrust in downward direction, and thus
          it supports the weight of the animal by reaction [2]. The
          narrow, orderly jet produced by the flexible foil in still
          fluid during the present experiments is accompanied by a
          corresponding thrust. In a sense, this is a type of
          hovering, which is simple and different from the known
          hovering mechanisms present in the birds and insects.

          The video showcasing all these flow-features is submitted to
          the \textit{Gallery of Fluid Motion 2010} which is annual
          showcase of fluid dynamics videos. This video can be seen at
          the following URL:

\href{http://ecommons.library.cornell.edu/bitstream/1813/8237/2/FLEXIBLE_FOIL_JET.mpg}{Video}. \\ \\

\textbf{\Large{References}} \\

          [1] T. Maxworthy. \textit{The fluid dynamics of insect
          flight.} Annu. Rev. Fluid Mech., 13, 329--350, 1981. \\

          [2] M. J. Lighthill. \textit{On the Weis-Fogh mechanism of
          lift generation.} J. Fluid Mech., 60(1), 1--17, 1973.
\end{document}